\documentclass[twocolumn,showpacs,preprintnumbers,amsmath,amssymb]{revtex4}

\begin{document}

\title{Why is Schrodinger's Equation Linear?}

\author{Rajesh R. Parwani}
\email{parwani@nus.edu.sg}

\affiliation{University Scholars Programme, National University of Singapore, 05-25 Block ADM, Kent Ridge, Singapore.}

\date{04 September 2004. Revised 10 December 2004}

\begin{abstract}
Information-theoretic arguments are used to obtain a link between the accurate linearity of Schrodinger's equation and Lorentz invariance: A possible violation of the latter at short distances would imply the appearance of nonlinear corrections to quantum theory. Nonlinear corrections can also appear in a Lorentz invariant theory in the form of higher derivative terms that are determined by a length scale, possibly the Planck length. It is suggested that the best place to look for evidence of such quantum nonlinear effects is in neutrino physics and cosmology.

\end{abstract}

\maketitle


\section{Quantum and Information Theories: A Review}

Many authors have pondered over the linearity of Schrodinger's equation, see for example \cite{nonlin,kibble}, and although various nonlinear modifications have been suggested, there has been no direct experimental evidence but only tiny upper bounds on the size of the corrections. Thus the puzzle really is the small magnitude of the potential nonlinearities: What sets the scale? 

If one subscribes to the philosophy that the laws of physics should be constructed so as to provide the most economical and unbiased representation of empirical facts, then the principle of maximum uncertainty \cite{jay} is the natural avenue by which to investigate Schrodinger's equation \cite{reg1,raj2} and its possible generalisations \cite{raj1}. 

Let me first briefly review the procedure discussed more fully in Refs.\cite{reg1, raj2}. To fix the notation, consider 
the Schrodinger equation for $N$ particles in $d+1$ dimensions, 
\begin{equation}
i\hbar \dot{\psi} = \left[ - {\hbar^2 \over 2} g_{ij} \partial_i \partial_j + V \right] \psi \, \label{schmulp}
\end{equation}
where $i,j= 1,2,......,dN$ and the configuration space metric is defined as $g_{ij} = \delta_{ij} /m_{(i)}$ with the symbol $(i)$ defined as   the smallest integer $\ge i/d$. That is, $i=1,...d$, refer to the Cartesian coordinates of the first particle of mass $m_1$, $i=d+1,.....2d$, to those of the second particle of mass $m_2$ and so on. The summation convention is used unless otherwise stated. 

It is useful to write the Schrodinger equation in a form which allows comparison with classical physics. The transformation $\psi = \sqrt{p} \ e^{iS / \hbar}$ decomposes the Schrodinger equation into two real equations,

\begin{eqnarray}
\dot{S} + {g_{ij} \over 2} \partial_iS \partial_j S + V -{{\hbar}^2  \over 2 } {g_{ij} \over \sqrt{p}} \ \partial_i \partial_j \sqrt{p} &=& 0 \, ,  \label{hj3} \\ 
\dot{p}  + g_{ij} \ \partial_i \left( p \partial_j S \right)  &=& 0 \, .  \label{cont3} 
\end{eqnarray}
The first equation is a generalisation of the usual Hamilton-Jacobi equation, the term with explicit $\hbar$ dependence   summarising the peculiar aspects of quantum theory. The second equation is the 
continuity equation expressing the conservation of probability. 
These equations may be obtained from a variational principle \cite{reg1}: one minimises the action

\begin{equation}
\Phi = \int  p \left[ \dot{S}  + {g_{ij} \over 2} \partial_i S \partial_j S   + V \right] dx^{Nd}  dt   \ + {{\hbar}^2 \over 8}   I_F \, \label{varmulp} 
\end{equation}
with respect to the variables $p$ and $S$. The positive quantity
\begin{equation}
I_F  \equiv  \int dx^{Nd}  dt \  g_{ij} \ p  \left({\partial_i p \over p} \right)  \left({\partial_j p \over p} \right) \,  \label{fish}
\end{equation}
is essentially the ``Fisher information" \cite{fisher, Kullback}. Since a broader probability distribution $p(x)$ represents a greater uncertainty in $x$, so $I_F$ is actually an inverse uncertainty measure.

The equations (\ref{varmulp}, \ref{fish}) were first used in Ref.\cite{reg1} to derive Schrodinger's equation as follows.  It is noted that without the term $I_F$, variation of Eq.(\ref{varmulp}) gives rise to equations describing a classical ensemble. Then one adopts the principle of maximum uncertainty \cite{jay} to constrain the probability distribution $p(x)$ characterising the ensemble: since it is supposed to represent some fluctuations of unknown origin, we would like to be as unbiased as possible in its choice. The constraint  is implemented in (\ref{varmulp}) by minimising $I_F$ when varying the classical action: $ {\hbar}^2 /8 $ is the Lagrange multiplier.

The work of \cite{reg1} was extended in \cite{raj2} in two ways: First, constraints that a suitable (for inferring quantum theory) information measure  should satisfy were made explicit. Then, the relevant measure was {\it constructed} from the physical constraints rather than postulated, thus motivating the structure of the linear Schrodinger equation: 
In brief, consider the same classical ensemble as in (\ref{varmulp}), but now constrained by a general (unknown) information measure $I$ and a lagrange multiplier $\lambda$. The six constraints used in \cite{raj2} were: positivity of $I$, locality, homogeniety, separability, Gallilean invariance, and the absence of higher number of derivatives (beyond second) in any product of terms in the action. The last condition will be abbreviated as ``AHD".  

Except for positivity (required for a sensible interpretation of $I$ as an inverse uncertainty measure), {\it the other conditions are already satisfied by the classical part of the action so one is not imposing anything new}. The locaility, separability and homogeneity constraints also have a natural physical interpretation within the context of the resulting quantum theory \cite{raj2}. 

The unique solution of the above conditions was shown \cite{raj2} to be precisely the measure $I_F$.  The Lagrange multiplier $\lambda$ must then have the dimension of (action)$^2$ thereby introducing the Planck constant into the picture; the equation of motion is then the {\it linear} Schrodinger equation. The AHD condition can also be given a physical interpretation \cite{raj2}: It means that other than a single Lagrange multiplier related to Planck's constant $\hbar$, no other parameter is introduced in the approach. {\it Thus within the information theoretic approach, the linear Schrodinger equation is the unique one-parameter extension of the classical dynamics.} For more details, please refer to \cite{raj2}.

\section{Nonlinearities}

The only conditions, among those metioned above, that can be relaxed without causing a drastic change in the usual physical  interpretation of the wavefunction are Gallilean invariance and AHD. Also, since only the rotational invariance part of Gallilean symmetry was used explicitly \cite{raj2},  one deduces that {\it within the information theory context and with the other conditions fixed}, it is rotational invariance and AHD which are responsible for the linearity of the Schrodinger equation.

Write $I= I_{F} + (I-I_{F})$. Now since $I_{F}$ is rotationally symmetric, satisfies AHD and is also the unique measure responsible for the linear theory, one concludes that within the above-mentioned context, the violation of rotational invariance {\it or} AHD is a necessary and sufficient condition for a nonlinear Schrodinger equation. {Of course once rotational invariance is broken there is no reason to continue using the classical metric $g_{ij}$ in the information measure.} 

Let us consider first the breaking of the AHD condition. Then higher derivatives appear and this implies, on dimensional grounds, the appearance of a new length scale. It is tempting to associate such a scale with Planck length and thus the effect of gravity though other possibilities exist \cite{raj1}. Thus in this scenario one would have a Lorentz invariant but nonlinear correction to quantum mechanics, with a new length scale determining the size of the nonlinear corrections.
 
Consider next the scenario whereby rotational invariance is broken in the non-relativistic quantum theory. Then Lorentz invariance should be broken in a relativistic version. The explicit form of the symmetry breaking nonlinear Schrodinger equation of course will depend on the relevant measure that is used. The simplest possibilty is to use information measures that are commonly adopted in statistical mechanics as this would provide a link with the maximum entropy method used in that field \cite{jay,Kullback}. Such measures were studied in \cite{raj1} and they lead unavoidably to the appearance of higher derivatives and a length scale that quantifies the symmetry breaking.

It is interesting to note that both scenarios of nonlinear quantum dynamics typically involve higher derivatives and a new length scale.  
 
There have been several proposals of nonlinear Schrodinger equations in the literature, see for example \cite{nonlin, kibble,doeb,aub} and references therein. However those studies were not conducted within an information theoretic framework and so those  equations often do not satisfy the equivalent of one or more of the  conditions used in \cite{raj2}. Or, when the nonlinear equations so constructed are local, homogeneous, separable and Gallilean invariant, yet they do not follow from a local variational action with a positive definite $I$. Thus there is no contradiction between the result of this paper or \cite{raj2}, and other studies of nonlinear quantum theory.

\section{Conclusion}
Within the information theoretic context, it has been established here that ``minimal" nonlinear extensions of Schrodinger's equation are associated either with the breaking of Lorentz symmetry or the presence of higher order derivatives.
By minimal I mean that the locality, separability and homogeneity required for the usual interpretation of the wavefunction are preserved. 

More specifically, the breaking of Lorentz invariance implies the appearance of nonlinear corrections to quantum theory and this has now been established without the use of specific examples of symmetry breaking information measures that were originally used in \cite{raj1}. 

Since empirical evidence suggests that Lorentz symmetry violation (if any) is expected to be very small, this then would explain the tiny size of potential nonlinearities. One may rephrase this using the concept of \cite{hooft}: the smallness of the nonlinearities would be ``natural" because in the limit of vanishing nonlinearity one would obtain a realisation of the full Lorentz symmetry.

In the alternate scenario, if Lorentz symmetry remains exact down to small distances, then any potential nonlinearity scale is set by a  length parameter, possibly the Planck length, and thus the smallness of the potential nonlinear corrections to Schrodingers may  be attributed to the weakness of gravity. 

Phenomenological consequences of a nonlinear symmetry breaking quantum dynamics have  been discussed in \cite{raj1}. However almost all the effects discussed there had more to do with the quantum nonlinearity than the symmetry breaking. Thus here I would like to re-highlight the suggestion that three puzzles: neutrino oscillations, dark energy and dark matter, might be  common manifestations of a nonlinear quantum theory. For example, the effect of quantum nonlinearities would be to give particles a contribution to their mass which  varies with energy \cite{raj1}. Since the varying component is tiny, it would be more apparent for neutrinos and may be responsible for their oscillations. Furthermore, while the effect of quantum nonlinearities on a single particle is small, the cumulative effect in very large systems, say of cosmological size, might be apprecaible.

Possible Lorentz violation has been a subject of study by many authors. The exciting possibility discussed in \cite{raj1} is its link with quantum nonlinearities. However it is technically much easier to study one of the corrections (e.g. nonlinear quantum theory) while keeping the other unchanged. One may view this approximation as an effective approach whereby the role of quantum nonlinearity in the abovementioned puzzles is studied within a Lorentz invariant interpretative framework \cite{raj4}.  On the otherhand, as discussed above, even in a Lorentz invariant theory one could have nonlinear corrections to quantum theory, so the effective approach covers both possibilities. \\

\section{Acknowledgments}

I am grateful to Prof. H. Elze for inviting me to participate in the stimulating workshop. I also thank all the organisers for their hospitality and Marcel Reginatto for several useful discussions.

\bibliography{apssamp}

\end{document}